\documentclass[12pt]{article}
\usepackage{amsmath}
\usepackage{graphicx}
\usepackage{enumerate}
\usepackage{natbib}
\usepackage{booktabs, caption, makecell}
\usepackage{subcaption}
\usepackage{multirow}
\usepackage{bm}
\usepackage{authblk}
\usepackage{xcolor}
\usepackage{colortbl}
\usepackage{tikz}
\usetikzlibrary{shapes.geometric, arrows.meta, positioning, fit}

\tikzset{
  startstop/.style={rectangle, rounded corners, minimum width=2.5cm, minimum height=1cm, draw},
  process/.style={rectangle, minimum width=2.5cm, minimum height=1cm, draw},
  decision/.style={diamond, aspect=2, minimum width=2cm, minimum height=1cm, draw},
  arrow/.style={thick, ->, >=Stealth},
redbox/.style={draw=black, thick, rounded corners, fit=#1, inner sep=2mm}
}

\usepackage{amssymb}

\newcommand*\squared[1]{\tikz[baseline=(char.base)]{
            \node[shape=rectangle,draw,inner sep=2pt, thick] (char) {#1};}}

\usepackage{enumitem}
\setlist[itemize]{leftmargin=*}

\usepackage{stackengine}

\newcommand{\blind}{1}

\begin{document}

\def\spacingset#1{\renewcommand{\baselinestretch}%
{#1}\small\normalsize} \spacingset{1}

%%%%%%%%%%%%%%%%%%%%%%%%%%%%%%%%%%%%%%%%%%%%%%%%%%%%%%%%%%%%%%%%%%%%%%%%%%%%%%

\if1\blind
{
  \title{\bf Two-level D- and A-optimal designs of Ehlich type with run sizes three more than a multiple of four}
  \author{Mohammed Saif Ismail Hameed$^1$}
    \author{Eric D. Schoen$^{1}$}
    \author{José Núñez Ares$^2$}
    \author{Peter Goos$^{1}$}
    \affil{$^1$Department of Biosystems, KU Leuven, Leuven, Belgium}
    \affil{$^2$EFFEX, Leuven, Belgium}
    %   \author{Author 1\thanks{
    % The authors gratefully acknowledge \textit{please remember to list all relevant funding sources in the unblinded version}}\hspace{.2cm}\\
    % Department of YYY, University of XXX\\
    % and \\
    % Author 2 \\
    % Department of ZZZ, University of WWW}
  \maketitle
} \fi

\if0\blind
{
  \bigskip
  \bigskip
  \bigskip
  \begin{center}
    {\LARGE\bf Two-level D- and A-optimal main-effects designs with run sizes three more than a multiple of four}
\end{center}
  \medskip
} \fi

\bigskip
\begin{abstract}
\noindent For the majority of run sizes $N$ where $N \leq 20$, the literature reports the best D- and A-optimal designs for the main-effects model which sequentially minimizes the aliasing between main effects and interaction effects and among interaction effects. The only series of run sizes for which all the minimally aliased D- and A-optimal main-effects designs remain unknown are those with run sizes three more than a multiple of four. To address this, in our paper, we propose an algorithm to generate all non-isomorphic D- and A-optimal main-effects designs for run sizes three more than a multiple of four. We enumerate all such designs for run sizes up to 19, report the numbers of designs we obtained, and identify those that minimize the aliasing between main effects and interaction effects and among interaction effects.
\end{abstract}

\noindent%
\vfill

\newpage
\spacingset{1.9} % DON'T change the spacing!

\section{Introduction} \label{sec:intro}

% As described earlier in Section (intro), the (first step) in constructing a D- and/or A-optimal design, is to first identify the potential candidate for its information matrix from the class of matrices $M$. 

% \noindent Therefore the problem of identifying the suitable candidate matrix (or matrices) in $M$ that may correspond to a D- and/or A-optimal design can be translated to identifying the optimal value(s) for $s$. 

To study the effects of a given number of factors on one or more responses, a common strategy is to choose an experimental design that optimizes a criterion that directly measures the information content of the experiment. One such criterion is the determinant of the information matrix $\mathbf{X'X}$, where $\mathbf{X}$ is the $N \times p$ model matrix corresponding to a linear regression model, $N$ is the number of experimental runs and $p$ is the total number of parameters. A design that maximizes this determinant is labeled D-optimal, where ‘D’ stands for determinant \citep{goos2011optimal}. D-optimal designs minimize the generalized variance of the estimates of the parameters in the regression model, as well as the volume of the parameters' confidence ellipsoid \citep{atkinson2007optimum}. An alternative design selection criterion is the trace of $(\mathbf{X'X})^{-1}$. A-optimal designs minimize this trace and consequently have the lowest average variance for all estimates of the model parameters \citep{atkinson2007optimum}. In the name A-optimal, ‘A’ stands for average.

To study main effects and interactions of quantitative factors, it is sufficient to only use two levels ($-1$ and $1$) for each factor in the design. Therefore, our work focuses on two-level D- and A-optimal designs. Given the importance of experimental factors’ main effects and the possible importance of their interactions, it would be ideal if an experimental plan were D- and A-optimal for the complete main-effects-plus-interactions model. However, for $k$ factors, D- and A-optimal experimental plans for this model require at least $1+ k+\binom{k}{2}$ runs, so that they are often too costly to be used in practice. For this reason, it is common to optimize the experimental design for the main effects, while making sure that the bias in the main-effect estimates is minimized and the information on the interactions is maximized. In this article, we focus on D- and A-optimal designs for the main-effects model, so that the model matrix $\mathbf{X}$ includes a column for the intercept and the remaining columns correspond to the $k$ factors and $p=1+k$. In our paper, we refer to the matrix $\mathbf{X}$ as the design. 

For any given pair of $N$ and $p$, many different D- and A-optimal designs exist. Some of these D- and A-optimal designs are isomorphic to each other, because it is possible to obtain one design from another by permuting its factor columns, permuting its rows, and/or switching the signs of the levels of one or more of its factors. All the designs that are isomorphic to each other form an isomorphism class. Designs of the same isomorphism class possess the same statistical qualities. Therefore, it makes sense to study only sets of non-isomorphic D- and A-optimal designs. Secondary criteria can then be applied to further distinguish between these designs. Because of the potential importance of two-factor interactions, two obvious secondary criteria are the aliasing between main effects and two-factor interactions and the aliasing between two-factor interactions, respectively.

A substantial body of literature has been devoted to complete catalogs of non-isomorphic D- and A-optimal two-level designs for small run sizes and to the identification of the D- and A-optimal designs that minimize the aliasing between main effects and two-factor interactions and among two-factor interactions. For cases where $N\equiv 0$ (mod 4), \citet{sun2008algorithmic} propose an algorithm to construct complete sets of non-isomorphic D- and A-optimal designs and report the designs that minimize aliasing between the main effects and interactions and among interactions for $N \leq 20$. Similarly, \citet{schoen2017two} generate minimally aliased D- and A-optimal designs for run sizes up to 36. \citet{hameed2025_m12} provide a construction method to generate complete sets of non-isomorphic D- and A-optimal designs for cases where $N\equiv 1$ (mod 4) and $N\equiv 2$ (mod 4), and report the minimally aliased D- and A-optimal designs for $N \leq 18$. For cases where $N\equiv 3$ (mod 4), which is known to be the most challenging case \citep{galil1980d}, no single construction method can generate complete sets of D- and A-optimal designs. Hence, the minimally aliased D- and A-optimal designs for such run sizes remain unknown. The most extensive results for $N\equiv 3$ (mod 4) can be found in \citet{hameed2025_dropping}. These authors study a complete collection of non-isomorphic D-optimal designs where $N \leq 19$ and $N \geq 2p - 5$ and report the minimally aliased designs within this collection. All remaining cases where $N\equiv 3$ (mod 4) are unexplored territory when it comes to generating non-isomorphic D- and A-optimal two-level designs. Although there exist methods that can construct a D- and A-optimal design, not a single method can generate the complete set of all non-isomorphic D- and A-optimal designs.

As the philosophy of optimal experimental design is to tailor the experimental plan to the situation at hand, there is a clear need for optimal experimental designs for any run size, especially considering the fact that more runs always lead to better standard errors, powers and prediction variances. Therefore, it is of crucial importance to study D- and A-optimal designs for $N\equiv 3$ (mod 4) in as much detail as the other three kinds of run sizes. In this article, we fill this gap by proposing an algorithm to construct complete sets of D- and A-optimal designs for $N\equiv 3$ (mod 4). For $N \leq 19$, we also report the minimally aliased D- and A-optimal designs for each combination of $N$ and $p$. Our work supersedes the work of \citet{hameed2025_dropping}, as their work only covers complete sets of D-optimal designs for cases where $N \geq 2p - 5$, whereas our method can be used to generate both D- and A-optimal designs for any combination of $N$ and $p$, provided $N \geq p$.

The remainder of this article is organized as follows. In Section \ref{sec:literature}, we provide a literature review of all the theoretical results relevant to our work. In Section \ref{ssec:our_work}, with reference to the existing literature, we detail the contributions we make in this article. In Section \ref{sec:algorithm}, we describe our enumeration algorithm to generate D- and/or A-optimal designs. In Section \ref{sec:enu_results}, we report the results of our enumeration and characterize the minimally aliased D- and A-optimal designs we obtained. Finally, in Section \ref{sec:conclusion}, we provide our conclusions.

\section{Literature review} \label{sec:literature} 

Much of the literature on D- or A-optimal two-level designs is concerned with the exact form of the information matrix $\mathbf{X'X}$. In case $N\equiv 0$ (mod 4), a design $\mathbf{X}$ is both D- and A-optimal if $\mathbf{X'X}=N\mathbf{I}_p$, where $\mathbf{I}_p$ is the $p$-dimensional identity matrix. The algorithm of \citet{sun2008algorithmic} uses this information to generate complete sets of non-isomorphic D- and A-optimal designs for $N \in \{12,16,20\}$. In case $N\equiv 1$ (mod 4), the information matrix for a D- and A-optimal design is $(N-1)\mathbf{I}_{p} + \mathbf{J}_{p}$, where $\mathbf{J}_{p}$ is a $p \times p$ matrix with all entries equal to 1 \citep{cheng1980optimality}. In case $N\equiv 2$ (mod 4), the information matrix $\mathbf{X'X}$ for a D- and A-optimal design is a block-diagonal matrix $\text{diag}(\mathbf{B}_{i}, \mathbf{B}_{j})$, where $\mathbf{B}_{i}=(N-2)\mathbf{I}_{i} + 2\mathbf{J}_{i}$, $\mathbf{B}_{j}=(N-2)\mathbf{I}_{j} + 2\mathbf{J}_{j}$, $i$ is the integer part of $p/2$, and $j = p - i$ \citep{jacroux1983optimality}. \citet{hameed2025_m12} use the results for $N\equiv 1$ (mod 4) and $N\equiv 2$ (mod 4) to construct all non-isomorphic designs for $N \in \{5,6,9,10,13,14,17,18\}$.

%of the form
%\begin{align} \label{form_n2}
%    \begin{bmatrix}
%        \mathbf{B}_{i} & \mathbf{0}_{ij} \\
%        \mathbf{0}_{ji} & \mathbf{B}_{j} 
%    \end{bmatrix}=\text{diag}(\mathbf{B}_{i}, \mathbf{B}_{j}),
%\end{align}

For $N\equiv 3$ (mod 4), which is the type of run size we focus on in this paper, the forms of the information matrices for D- or A-optimal designs are more complex and more variable. Generally, they belong to a class of matrices defined by \citet{Ehlich1964}. These matrices involve $s$ blocks of the form $(N-3)\mathbf{I}_{r_i} + 3\mathbf{J}_{r_i}$ (where $i=1,\dots,s$) on the diagonal and entries of $-1$  elsewhere. We introduce the Ehlich matrices in Section~\ref{ssec:m_n_p}. Next, in Section~\ref{ssec:m3_theoretical_results}, we discuss which Ehlich matrices have the potential to serve as information matrices for D- and/or A-optimal designs, and which therefore form a starting point for our enumeration algorithm. Finally, in Section~\ref{ssec:sufficiency}, we review the existing constructions of D- and A-optimal designs with information matrices of the Ehlich form.

\subsection{Ehlich matrices} \label{ssec:m_n_p}

An Ehlich matrix $\mathbf{K}_{(N,p,s)}$, where $N$, $p$ and $s$ are strictly positive integers and $N \geq p \geq s$, is a $p \times p$ matrix of the form
%Consider the following class of matrices from  \citet{Ehlich1964}. For a given pair ($N,p$) with $N \geq p$, let $M(N,p)$ correspond to the class of all $p \times p$ matrices partitioned as
\begin{align} \label{form_m3}
\begin{bmatrix}
\mathbf{B}_{11} & \mathbf{A}_{12} & \cdots & \mathbf{A}_{1s}\\
\mathbf{A}_{21} & \mathbf{B}_{22} & \cdots & \mathbf{A}_{2s}\\
\vdots & \vdots & \ddots & \vdots\\
\mathbf{A}_{s1} & \mathbf{A}_{s2} & \cdots & \mathbf{B}_{ss}
\end{bmatrix},
\end{align}
where each of the $s$ $\mathbf{B}_{ii}$ matrices is a block of the form $(N-3)\mathbf{I}_{r_i} + 3\mathbf{J}_{r_i}$ and every $\mathbf{A}_{ij}$ matrix is an $r_i \times r_j$ matrix whose entries are all $-1$. In the event $p$ is not a multiple of $s$, $\mathbf{K}_{(N,p,s)}$ involves $\mathbf{B}_{ii}$ matrices of two contiguous sizes, $r$ and $r+1$. We denote the numbers of $\mathbf{B}_{ii}$ matrices with sizes $r$ and $r+1$ by $u$ and $v$, respectively. In the event $p$ is a multiple of $s$, all $\mathbf{B}_{ii}$ matrices in $\mathbf{K}_{(N,p,s)}$ have the same size, such that $r_i=r=p/s$, $u=s$ and $v=0$. Obviously, $v=s-u$ and $p = ur + v(r+1) = sr + v$ holds true for both cases. Key features of any Ehlich matrices $\mathbf{K}_{(N,p,s)}$ are that all diagonal elements equal $N$ and that all off-diagonal elements are $3$ and/or $-1$.

%In the latter case, we denote the number of $\mathbf{B}_{ii}$ matrices of size $r$ bu $u$ and the number of $\mathbf{B}_{ii}$ matrices of size $r+1$ by $v$, where $v=s-u$.

%. The total number of blocks equals $s$.   Further, each matrix either has all blocks $\mathbf{B}_{ii}$ of the same size or two contiguous sizes such that $r_i = r \text{ for }i=1,\dots,u\;\text{ and }r_i = r+1\text{ for }i=u+1,\dots,s$. 

Now, for any given $N\equiv 3$ (mod 4), it is possible to construct an Ehlich matrix $\mathbf{K}_{(N,p,s)}$ for every $s \in \{1,\dots,p\}$. At one end of the spectrum, when $s=1$, we obtain the Ehlich matrix $\mathbf{K}_{(N,p,1)}$ in which all off-diagonal elements equal 3. At the other end of the spectrum, when $s=p$, we obtain the Ehlich matrix $\mathbf{K}_{(N,p,p)}$ in which all off-diagonal elements are $-1$. We denote the complete set of $p$ Ehlich matrices that can be constructed for a given $N$ by $M(N,p)$. Each matrix in $M(N,p)$ can be uniquely characterized by the number $s$, since $s$ uniquely determines the values for $r$, $u$ and $v$. In general, when $p$ is an integer multiple of $s$, $r=p/s$, $u=s$ and $v=0$. For instance, when $s=1$, $r=p$, $u=1$ and $v=0$, and when $s=p$, $r=1$, $u=p$ and $v=0$. For the majority of the Ehlich matrices, however, $p$ is not an integer multiple of $s$, in which case $r$ is equal to the integer part of the fraction $p/s$, $u=s(r+1)-p$ and $v=s-u$. An example of such an Ehlich matrix is the matrix $\mathbf{K}_{(15,14,4)}$, where $s=4$ (hence, $r=3$ and $u=v=2$. That matrix has $u=2$ blocks of size $r=3$, $\mathbf{B}_{11}$ and $\mathbf{B}_{22}$, and $v=2$ blocks of size $r+1=4$, $\mathbf{B}_{33}$ and $\mathbf{B}_{44}$, and is given below:

%can be distinguished by the set $\{s,r,u,v\}$ where $u$ is the number of blocks of size $r$, $v$ is the number of blocks of size $r+1$ and $s$ is the total number of blocks so that $s=u+v$ and $p = ur + v(r+1) = sr + v$. However, the values for $p$ and $s$ determine the values for the other three variables $(r,u,v)$, as follows:
%\begin{itemize}
 %   \item When $p$ is not divisible by $s$, the matrix will have some blocks of size $r$ followed by some blocks of size $r+1$ (i.e $u,v > 0$). In this case, $r$ is the integer part of the fraction $p/s$, $u = s(r+1)-p$ and $v=s-u$.
  %  \item When $p$ is divisible by $s$, the matrix has all of its blocks of the same size. In this case, there are two sets of values for $(r,u,v)$: (i) $r=p/s$, $u=s$ and $v=0$, and (ii) $r=(p/s)-1$, $u=0$ and $v=s$. Note that both sets of values for $(r,u,v)$ virtually represent the same matrix in $M(N,p)$. In this paper, we use the representation (i) to refer to Ehlich matrices for cases where $p$ is divisible by $s$. 
%\end{itemize}

%\noindent Obviously, $s$ must be equal to or smaller than $p$. Hence, for a given combination $N$ and $p$, there are exactly $p$ matrices in $M(N,p)$ identified by all possible values for $s$ in the set $\{1,2,\dots,p \}$. Therefore, every Ehlich matrix can be identified using $N, p$ and $s$. We denote any such matrix as $\mathbf{K}_{(N,p,s)}$. Note that all values on the diagonal equal $N$, and all off-diagonal values are either $-$1 or 3.

%As an illustration, Equation (\ref{eq:15_14_4}) shows the matrix $\mathbf{K}_{(15,14,4)}$ where $(r,u,v) = (3,2,2)$.

\spacingset{1}
\begin{equation} \label{eq:15_14_4}
\mathbf{K}_{(15,14,4)} = \left[
\begin{array}{*{14}{r}} % 14 right-aligned columns
15 & \phantom{0}3 & \phantom{0}3 & -1 & -1 & -1 & -1 & -1 & -1 & -1 & -1 & -1 & -1 & -1 \\
\phantom{0}3 & 15 & \phantom{0}3 & -1 & -1 & -1 & -1 & -1 & -1 & -1 & -1 & -1 & -1 & -1 \\
\phantom{0}3 & \phantom{0}3 & 15 & -1 & -1 & -1 & -1 & -1 & -1 & -1 & -1 & -1 & -1 & -1 \\
-1 & -1 & -1 & 15 & \phantom{0}3 & \phantom{0}3 & -1 & -1 & -1 & -1 & -1 & -1 & -1 & -1 \\
-1 & -1 & -1 & \phantom{0}3 & 15 & \phantom{0}3 & -1 & -1 & -1 & -1 & -1 & -1 & -1 & -1 \\
-1 & -1 & -1 & \phantom{0}3 & \phantom{0}3 & 15 & -1 & -1 & -1 & -1 & -1 & -1 & -1 & -1 \\
-1 & -1 & -1 & -1 & -1 & -1 & 15 & \phantom{0}3 & \phantom{0}3 & \phantom{0}3 & -1 & -1 & -1 & -1 \\
-1 & -1 & -1 & -1 & -1 & -1 & \phantom{0}3 & 15 & \phantom{0}3 & \phantom{0}3 & -1 & -1 & -1 & -1\\
-1 & -1 & -1 & -1 & -1 & -1 & \phantom{0}3 & \phantom{0}3 & 15 & \phantom{0}3 & -1 & -1 & -1 & -1\\
-1 & -1 & -1 & -1 & -1 & -1 & \phantom{0}3 & \phantom{0}3 & \phantom{0}3 & 15 & -1 & -1 & -1 & -1\\
-1 & -1 & -1 & -1 & -1 & -1 & -1 & -1 & -1 & -1 & 15 & \phantom{0}3 & \phantom{0}3 & \phantom{0}3\\
-1 & -1 & -1 & -1 & -1 & -1 & -1 & -1 & -1 & -1 & \phantom{0}3 & 15 & \phantom{0}3 & \phantom{0}3\\
-1 & -1 & -1 & -1 & -1 & -1 & -1 & -1 & -1 & -1 & \phantom{0}3 & \phantom{0}3 & 15 & \phantom{0}3\\
-1 & -1 & -1 & -1 & -1 & -1 & -1 & -1 & -1 & -1 & \phantom{0}3 & \phantom{0}3 & \phantom{0}3 & 15\\
\end{array}
\right]
\end{equation}\\

\spacingset{1.9}
%Clearly, the initial two diagonal blocks are sized $3 \times 3$, while the remaining two blocks are sized $4 \times 4$. 

\citet{galil1980d} show that the determinant of an Ehlich matrix $\textbf{K}_{(N,p,s)}$ equals
\begin{equation*}
    (N-3)^{p-s} \bigg( 1-\sum_{i=1}^{s}\frac{r_i}{L_i} \bigg) \prod_{i=1}^{s}L_i,
\end{equation*}
while \citet{sathe1989optimal} show that the trace of its inverse equals
\begin{equation*}
    \sum_{i=1}^{s}L_i^{-1} + \frac{p-s}{N-3} + \sum_{i=1}^{s}\frac{r_i}{{L_i}^2} \bigg/\bigg(1-\sum_{i=1}^{s}\frac{r_i}{L_i}\bigg).
\end{equation*}
In these expressions, $L_i = N-3+4r_i$. 

\subsection{Connection with D- and A-optimality} \label{ssec:m3_theoretical_results}

The class of Ehlich matrices is helpful in proving the D- and A-optimality of two-level designs $\mathbf{X}$ with $N$ rows and $p$ columns. This is due to two technical results. The first of these results, due to \citet{Ehlich1964} and \citet{galil1980d}, is that
\begin{equation} \label{eq:ehlich_condition}
    \text{max\{det}(\mathbf{X'X}): \mathbf{X} \in D(N,p)\} \leq \text{max\{det}(\mathbf{K}_{(N,p,s)}): \mathbf{K}_{(N,p,s)} \in M(N,p)\}, 
\end{equation}
where $D(N,p)$ represents the set of all possible $N \times p$ two-level $\mathbf{X}$ matrices. The second of these results, due to \citet{sathe1989optimal}, is that
\begin{equation} \label{eq:sathe_condition}
    \text{min\{tr}(\mathbf{X'X})^{-1}: \mathbf{X} \in D(N,p)\} \geq \text{min\{tr}(\mathbf{K}_{(N,p,s)}^{-1}): \mathbf{K}_{(N,p,s)} \in M(N,p)\}.
\end{equation}

The first result implies that, if we manage to find an $N \times p$ two-level matrix $\mathbf{X}$ for which $\text{det}(\mathbf{X'X})$ is equal to the maximum determinant that can be reached by any Ehlich matrix $\mathbf{K}_{(N,p,s)} \in M(N,p)$, then that matrix $\mathbf{X}$ corresponds to a D-optimal design. The second result implies that, if we manage to find an $N \times p$ two-level matrix $\mathbf{X}$ for which $\text{tr}(\mathbf{X'X})^{-1}$ is equal to the minimum trace that can be reached by the inverse of any Ehlich matrix $\mathbf{K}_{(N,p,s)} \in M(N,p)$, then that matrix $\mathbf{X}$ corresponds to an A-optimal design. \citet{Ehlich1964} and \citet{galil1980d} used the first result to construct some D-optimal designs, while \citet{sathe1989optimal} used the second result to construct some two-level A-optimal designs.

In the previous section, we presented the standard representation of the Ehlich matrices. However, any permutation of the rows and the corresponding permutation of the columns also results in a matrix with the same determinant and the same trace of its inverse. Hence, any design that produces the same determinant (or trace of the inverse) for its information matrix as that of a given Ehlich matrix $\mathbf{K}_{(N,p,s)}$, and whose information matrix is not identical but only a permuted form of the said Ehlich matrix, implies that it is possible to permute the columns of the design such that its new information matrix is identical to the standard form given in Equation (\ref{form_m3}). The proposed enumeration uses this fact to employ a different ordering of the rows and columns in the design during enumeration than the one given in Equation (\ref{form_m3}).

%\noindent The maximum value for the determinant that can be achieved by the information matrix of any two-level design with $N$ runs and $p$ columns is thus always smaller than or equal to the determinant of a matrix that attains the largest value for its determinant among all $p$ matrices in $M(N,p)$. 

%\citet{Ehlich1964} used the first result to identify potential information matrices for two-level D-optimal designs with $N=p$, while \citet{galil1980d} used it to construct some D-optimal designs for several cases where $N > p$. \citet{sathe1989optimal} used the second result to construct some two-level A-optimal designs.

The most relevant Ehlich matrices $\mathbf{K}_{(N,p,s)} \in M(N,p)$ are those with the maximum determinant and those that have the smallest trace of their inverse. These matrices are potential information matrices of an $N$-run two-level D- and/or A-optimal design with $p$ columns. We used the adjective potential because, for some of these Ehlich matrices, there exists no $N$-run two-level design for which $\mathbf{X'X}=\mathbf{K}_{(N,p,s)}$. For many combinations of $N$ and $p$, however, multiple $\mathbf{X}$ matrices exist with an information matrix of the optimal Ehlich form. 

Another issue is that, for some combinations of $N$ and $p$, more than one Ehlich matrix possesses the maximum determinant or minimizes the trace of the inverse. In that event, more than one type of D- or A-optimal design exists. Consider, for example, the case where $N=15$. Tables~\ref{subtab:d_eff_n15} and~\ref{subtab:a_eff_n15} shows the D- and A-efficiencies, respectively, for all matrices in $M(15,p)$ for $p \in \{4,\dots,15\}$  and $s \in \{1,\dots,p\}$. Any given column in these tables shows the efficiencies of all $p$ Ehlich matrices in $M(15,p)$, relative to the maximum determinant and relative to the minimum trace achievable by an Ehlich matrix within $M(15,p)$. In most columns, the maximum efficiency of 100\% is reached by only one Ehlich matrix. D-optimal combinations of $p$ and $s$ are bordered in Table~\ref{subtab:d_eff_n15}, while A-optimal combinations are shaded in Table~\ref{subtab:a_eff_n15}. 

\spacingset{1}
\begin{table}%[ht]
  \centering
  \scriptsize
  % \addtolength{\tabcolsep}{-0.6em}
  % \setlength\extrarowheight{5pt}
  \begin{subtable}{\linewidth}
  \caption{D-efficiencies}
  \addtolength{\tabcolsep}{-0.6em}
  % \addtolength{\tabcolsep}{-0.5em}
    \begin{tabular}{cccccccccccccc}
    \multirow{15}[1]{*}{($s$)} & \multicolumn{1}{r|}{15} &       &          &          &          &          &          &          &          &          &          &          & 93.89\% \\
          & \multicolumn{1}{r|}{14} &        &          &          &          &          &          &          &          &          &          & 96.96\%  & 95.61\%  \\
          & \multicolumn{1}{r|}{13} &       &          &          &          &          &          &          &          &          & 98.47\%  & 97.78\%  & 96.81\% \\
          & \multicolumn{1}{r|}{12} &       &          &          &          &          &          &          &          & 99.31\%  & 98.93\%  & 98.41\%  & 97.69\% \\
          & \multicolumn{1}{r|}{11} &       &          &          &          &          &          &          & 99.80\%  & 99.57\%  & 99.29\%  & 98.90\%  & 98.36\% \\
          & \multicolumn{1}{r|}{10} &       &          &          &          &          &          & \squared{100.00\%} & 99.92\%  & 99.75\%  & 99.56\%  & 99.27\%  & 98.88\% \\
          & \multicolumn{1}{r|}{9} &       &          &          &          &          & \squared{100.00\%} & \squared{100.00\%} & 99.98\%  & 99.88\%  & 99.76\%  & 99.56\%  & 99.28\% \\
          & \multicolumn{1}{r|}{8} &       &          &          &          & \squared{100.00\%} & 99.90\%  & 99.96\%  & \squared{100.00\%} & 99.96\%  & 99.90\%  & 99.78\%  & 99.59\% \\
          & \multicolumn{1}{r|}{7} &       &          &          & \squared{100.00\%} & 99.80\%  & 99.77\%  & 99.89\%  & 99.98\%  & \squared{100.00\%} & \squared{100.00\%} & 99.94\%  & 99.86\% \\
          & \multicolumn{1}{r|}{6} &       &          & \squared{100.00\%} & 99.70\%  & 99.58\%  & 99.61\%  & 99.78\%  & 99.93\%  & \squared{100.00\%} & \squared{100.00\%} & \squared{100.00\%} & \squared{100.00\%} \\
          & \multicolumn{1}{r|}{5} &       & \squared{100.00\%} & 99.58\%  & 99.37\%  & 99.33\%  & 99.42\%  & 99.65\%  & 99.72\%  & 99.76\%  & 99.84\%  & 99.91\%  & 99.98\% \\
          & \multicolumn{1}{r|}{4} & \squared{100.00\%} & 99.43\%  & 99.14\%  & 99.03\%  & 99.06\%  & 99.00\%  & 99.14\%  & 99.30\%  & 99.42\%  & 99.47\%  & 99.53\%  & 99.60\% \\
          & \multicolumn{1}{r|}{3} & 99.21\%  & 98.83\%  & 98.67\%  & 98.35\%  & 98.24\%  & 98.31\%  & 98.40\%  & 98.52\%  & 98.62\%  & 98.67\%  & 98.73\%  & 98.82\% \\
        & \multicolumn{1}{r|}{2} & 98.40\%  & 97.73\%  & 97.41\%  & 97.05\%  & 96.92\%  & 96.83\%  & 96.96\%  & 97.01\%  & 97.07\%  & 97.11\%  & 97.17\%  & 97.20\% \\
        & \multicolumn{1}{r|}{1} & 95.84\%  & 95.07\%  & 94.49\%  & 94.08\%  & 93.83\%  & 93.72\%  & 93.77\%  & 93.83\%  & 93.84\%  & 93.89\%  & 93.94\%  & 93.99\% \\
\cmidrule{3-14}    \multicolumn{2}{c}{\multirow{2}[1]{*}{}} & 4 & 5 & 6 & 7 & 8 & 9 & 10 & 11 & 12 & 13 & 14 & 15 \\
    \multicolumn{2}{c}{} & \multicolumn{12}{c}{($p$)} \\
    \end{tabular}% 
  \label{subtab:d_eff_n15}%
  \end{subtable}
  \begin{subtable}{\linewidth}
  \caption{A-efficiencies}
  \addtolength{\tabcolsep}{-0.35em}
  % \addtolength{\tabcolsep}{-0.5em}
    \begin{tabular}{cccccccccccccc}
    \multirow{15}[1]{*}{($s$)} & \multicolumn{1}{r|}{15} &       &          &          &          &          &          &          &          &          &          &          & 59.26\% \\
          & \multicolumn{1}{r|}{14} &        &          &          &          &          &          &          &          &          &          & 78.57\%  & 70.18\%  \\
          & \multicolumn{1}{r|}{13} &       &          &          &          &          &          &          &          &          & 87.87\%  & 83.90\%  & 77.97\% \\
          & \multicolumn{1}{r|}{12} &       &          &          &          &          &          &          &          & 93.14\%  & 90.96\%  & 87.99\%  & 83.71\% \\
          & \multicolumn{1}{r|}{11} &       &          &          &          &          &          &          & 96.33\%  & 95.06\%  & 93.39\%  & 91.16\%  & 88.03\% \\
          & \multicolumn{1}{r|}{10} &       &          &          &          &          &          & 98.29\%  & 97.54\%  & 96.57\%  & 95.31\%  & 93.63\%  & 91.32\% \\
          & \multicolumn{1}{r|}{9} &       &          &          &          &          & 99.50\%  & 98.99\%  & 98.46\%  & 97.75\%  & 96.81\%  & 95.56\%  & 93.86\% \\
          & \multicolumn{1}{r|}{8} &       &          &          &          & \cellcolor{blue!20}100.00\% & 99.82\%  & 99.49\%  & 99.14\%  & 98.65\%  & 97.97\%  & 97.06\%  & 95.81\% \\
          & \multicolumn{1}{r|}{7} &       &          &          & \cellcolor{blue!20}100.00\% & \cellcolor{blue!20}100.00\% & 99.98\%  & 99.81\%  & 99.63\%  & 99.32\%  & 98.86\%  & 98.21\%  & 97.78\% \\
          & \multicolumn{1}{r|}{6} &       &          & \cellcolor{blue!20}100.00\% & 99.70\%  & 99.87\%  & \cellcolor{blue!20}100.00\% & 99.97\%  & 99.94\%  & 99.79\%  & 99.62\%  & 99.45\%  & 99.27\% \\
          & \multicolumn{1}{r|}{5} &       & \cellcolor{blue!20}100.00\% & 99.38\%  & 99.28\%  & 99.62\%  & 99.89\%  & \cellcolor{blue!20}100.00\% & \cellcolor{blue!20}100.00\% & \cellcolor{blue!20}100.00\% & \cellcolor{blue!20}100.00\% & \cellcolor{blue!20}100.00\% & \cellcolor{blue!20}100.00\% \\
          & \multicolumn{1}{r|}{4} & \cellcolor{blue!20}100.00\% & 99.02\%  & 98.67\%  & 98.77\%  & 99.26\%  & 99.40\%  & 99.47\%  & 99.64\%  & 99.79\%  & 99.82\%  & 99.88\%  & 99.97\% \\
          & \multicolumn{1}{r|}{3} & 98.53\%  & 97.96\%  & 97.88\%  & 97.70\%  & 98.04\%  & 98.41\%  & 98.45\%  & 98.61\%  & 98.78\%  & 98.85\%  & 98.95\%  & 99.09\% \\
        & \multicolumn{1}{r|}{2} & 97.01\%  & 96.07\%  & 95.81\%  & 95.67\%  & 96.04\%  & 96.25\%  & 96.41\%  & 96.56\%  & 96.72\%  & 96.83\%  & 96.97\%  & 97.10\% \\
        & \multicolumn{1}{r|}{1} & 92.86\%  & 92.05\%  & 91.67\%  & 91.67\%  & 92.05\%  & 92.39\%  & 92.62\%  & 92.89\%  & 93.14\%  & 93.37\%  & 93.59\%  & 93.83\%  \\
\cmidrule{3-14}    \multicolumn{2}{c}{\multirow{2}[1]{*}{}} & 4 & 5 & 6 & 7 & 8 & 9 & 10 & 11 & 12 & 13 & 14 & 15 \\
    \multicolumn{2}{c}{} & \multicolumn{12}{c}{($p$)} \\
    \end{tabular}% 
  \label{subtab:a_eff_n15}%
  \end{subtable}
  \caption{D- and A-efficiencies for matrices $\mathbf{K}_{(15,p,s)}$ in $M(15,p)$ for different numbers of blocks ($s$) and parameters ($p$) for the main-effects model. In each column, the bordered and shaded cells indicate the matrices $\mathbf{K}_{(15,p,s)}$ with maximum D- and A-efficiencies, respectively.}
  \label{tab:eff_n15}
  \end{table}
\spacingset{1.9}

For several values of $p$, however, there are two optimal $s$ values, implying that there are two optimal Ehlich matrices. For instance, when $p=10$, both $s=10$ and $s=9$ yield a D-efficiency of 100\%, so that both $\mathbf{K}_{(15,10,10)}$ and $\mathbf{K}_{(15,10,9)}$ reach the maximum determinant. Therefore, both these matrices are candidate information matrices for D-optimal designs with $N=15$ and $p=10$. However, neither of these matrices is a candidate for an A-optimal design: Table~\ref{subtab:a_eff_n15} shows that an $s$ value of 5 is A-optimal for $N=15$ and $p=10$. So, of all matrices in $M(15,p)$, it is the matrix $\mathbf{K}_{(15,10,5)}$ which has the lowest trace of its inverse of all matrices in $M(15,10)$. Conversely, for $p=8$, there are two candidate information matrices for A-optimal designs and only one candidate information matrix for D-optimal designs.

Now, for a given pair of $(N,p)$, denote the set of all $s$ values that yield candidate information matrices for a D-optimal design by $S_{opt}^D$ and the set of all $s$ values that yield candidate information matrices for an A-optimal design by $S_{opt}^A$. \citet{gk_on_the_char} provide bounds for the values of $s$ in $S_{opt}^D$, and \citet{sathe1989optimal} derive similar results for $S_{opt}^A$. \citet{gk_on_the_char} further prove that $|S_{opt}^D| \in \{1,2\}$, and provide the exact conditions for which $|S_{opt}^D|=2$. Similarly, \citet{sathe1989optimal} prove that $|S_{opt}^A| \in \{1,2\}$ and give conditions for which $|S_{opt}^A|=2$. Thus, there are at most two D-optimal $s$ values and at most two A-optimal $s$ values. Obviously, when there are two optimal $s$ values, two distinct classes of D- or A-optimal designs may exist with a different information matrix of the Ehlich type. \citet{galil1980d}, \citet{sathe199912} and \citet{king2020direct} report the D-optimal $s$ values for many combinations of $N$ and $p$, where $p \leq N < 100$. Similarly, \citet{sathe1990construction} report A-optimal $s$ values.

%, provided both information matrices have a solution for $\mathbf{X}$ in the expression $\mathbf{X'X} = \mathbf{K}_{(N,p,s)}$. 
%the sets can be identical, mutually exclusive or overlapping.

\subsection{Gap in the literature} \label{ssec:sufficiency}

After identifying the optimal information matrices of D- or A-optimal designs, the corresponding $\mathbf{X}$ matrix should be searched for. When the optimal value for $s$ equals $p$, the design matrix is a two level orthogonal array with run size $N+1$ minus any one row \citep{hameed2025_dropping}. In other cases, more elaborate construction methods are necessary (see for instance, \citet{gk_opt_wei_des,gail1982construction,kounias1983some,sathe1990construction,sathe1991further,farmakis1991constructions}, and \citet{king2020direct}).

In several cases, a solution for $\mathbf{X}$ does not exist. For these cases, either one of the following assertions concerning the value of the determinant (or the trace of the inverse) of the specific matrix $\mathbf{K}_{(N,p,s)}$ is true: (i) this value cannot be attained by the information matrix of any two-level design, or (ii) this value can be attained by the information matrix of a two-level design that is not the same matrix as the optimal Ehlich matrix $\mathbf{K}_{(N,p,s)}$. For example, consider the case where $N=p=7$. For this case, the matrix $\mathbf{K}_{(7,7,5)}$ is the candidate information matrix for a D-optimal design. However, no corresponding solution for $\mathbf{X}$ exists \citep{galil1980d}. For the same case, $\mathbf{K}_{(7,7,4)} \in M(7,7)$ attains the smallest value of the trace of its inverse among all matrices $\mathbf{K}_{(7,7,s)} \in M(7,7)$. \citet{mood1946hotelling} shows that this matrix has a solution for $\mathbf{X}$ and that this solution is also D-optimal. The determinant of its information matrix is lower than the determinant of the original candidate information matrix $\mathbf{K}_{(7,7,5)}$.

These findings highlight that the results of \citet{Ehlich1964} and \citet{sathe1989optimal} are sufficient (and not necessary) conditions for the existence of optimal designs with an information matrix in the form of an Ehlich matrix. In addition, even if there exists a solution for $\mathbf{X}$ corresponding to an optimal Ehlich matrix, there may be other $\mathbf{X}$ matrices with the same D- or A-efficiency but with a information matrix not contained in $M(N,p)$. For some combinations of $N$ and $p$, the corresponding $\mathbf{K}_{(N,p,s)}$ matrix has been proven to be uniquely optimal whereas in other cases a proof is still missing \citep{galil1980d}. 
 
\section{Our contribution} \label{ssec:our_work}

To the best of our knowledge, there is no single construction method that can generate designs for all cases. Also, no single method can generate all non-isomorphic designs for any information matrix of the form $\mathbf{K}_{(N,p,s)}$, except when $p=s$ \citep{hameed2025_dropping}. This is because existing methods construct a single solution for $\mathbf{X}$ as opposed to all non-isomorphic solutions. By contrast, our design enumeration algorithm constructs all non-isomorphic $\mathbf{X}$ matrices that solve the expression $\mathbf{X'X} = \mathbf{K}_{(N,p,s)}$ (if a solution for $\mathbf{X}$ exists) for any combination of $N,p$ and $s$. We use our algorithm to construct complete sets of non-isomorphic D- and/or A-optimal designs for $N \in \{7,11,15,19\}$ and an information matrix of the form $\mathbf{K}_{(N,p,s)}$.

\section{Algorithm} \label{sec:algorithm}

Our enumeration procedure for designs with run size $N\equiv 3$ (mod 4) and an information matrix of the form $\mathbf{K}_{(N,p,s)}$ uses a multi-stage column extension procedure similar to that of \citet{hameed2025_m12} for run sizes $N\equiv 1$ (mod 4) and $N\equiv 2$ (mod 4). However, due to the more complex nature of the D- and A-optimal information matrices for run size $N\equiv 3$ (mod 4) is more involved. 
In Section \ref{ssec:m3_cc}, we detail the column extension step of our enumeration procedure. In Section \ref{ssec:m3_setcolumns}, we describe the set of columns we consider at each column extension step. Finally, in Section \ref{ssec:m3_esteps}, we discuss the isomorphism testing of the designs produced at each column extension step.

\subsection{Column extension} \label{ssec:m3_cc}

Just like other enumeration algorithms in the design of experiments literature  (see, for instance, \citet{chen1993catalogue,li2004design, schoen2010complete, ryan2010minimum}), our enumeration procedure uses a column-by-column extension. As the information matrix of every design to be produced by our enumeration procedure is an Ehlich matrix, the columns of every design should possess specific technical properties. We first discuss these properties, as this is essential for a good understanding of the different steps in our enumeration procedure. Next, we introduce the initial design that starts the enumeration for any given run size. Finally, we discuss the extension approach, departing from the initial design and taking the design columns' required properties into account.

\subsubsection{Required column properties} \label{sub:all_constraints}

% Taking the first column of the designs $\mathbf{X}$ to be the intercept column, the first row and column in the information matrix $\mathbf{K}_{(N,p,s)}$ tells us what the sum of each factor column in the design must be, and all off-diagonal values other than those in the first row and column give us the possible values for the inner products between any two factor columns. 

In the event we intend to enumerate designs with $N$ runs, $p-1$ factors (and thus $p$ columns in the model matrix {\bf X}) and a certain $s$ value, the corresponding Ehlich matrix $\mathbf{K}_{(N,p,s)}$ determines what columns are required in the model matrix corresponding to the designs. More specifically, the $p$ columns in the model matrix should be composed of $s$ groups of columns. The inner product between any pair of columns within a group should be $3$, while the inner product between any pair of columns belonging to different groups should be $-1$. The number of columns within any group is dictated by the values $u$, $v$ and $r$ corresponding to the $s$ value under consideration (see Section~\ref{ssec:m_n_p}). 

%To see this, recall that every Ehlich matrix $\mathbf{K}_{(N,p,s)}$ has specific $s$, $r$, $u$ and $v$ values (see Section~\ref{ssec:m_n_p}).

% For instance, when $p=s$, any design $\mathbf{X}$ of the form $\mathbf{K}_{(N,p,p)}$ must have all of its $p-1$ factor columns sum to $-1$ and the inner products between any pair of these $p-1$ factor columns must also equal $-1$. Consider the matrix $\mathbf{K}_{(7,5,5)}$ given in Equation (\ref{eq:7_5_5}). Any design of the form $\mathbf{K}_{(7,5,5)}$ must have all four of its factor columns sum to $-1$ and the inner products between any pair of these four factor columns must also equal $-1$. Therefore, to construct any design $\mathbf{X}$ of the form $\mathbf{K}_{(N,p,p)}$, we only need to consider columns that sum to $-1$.

%In fact the $p$ columns of the design matrix can be divided into $s$ groups, such that the inner product between any pair of columns within a group equals $3$, while the inner product between any pair of columns belonging to different groups equals $-1$. 

A special case arises when $s=p$. In that case, $r=1$, $u=s$ and $v=0$. In this case, all groups of columns are singletons, there are no pairs of columns whose inner product needs to be $3$, and the off-diagonal elements in the Ehlich matrix are all $-1$. This implies that the inner products between all the designs' factor columns all equal $-1$ and that all $p-1$ factor columns sum to $-1$ whenever $s=p$. Because all required columns are similar when $s=p$, enumerating designs is quite simple for that case. Whenever $s<p$, the enumeration is more involved.

When $r=1$ and $v>0$, the designs' model matrices involve $u$ groups of one column and $v$ groups of two columns. In that case, the columns in the first kind of group must have inner products equal to $-1$ with every other column in the design matrix. As the intercept column is in one of the groups, it must have an inner product of 3 with all other columns in its group (if any). This means that all factor columns in the group of the intercept must sum to 3, and hence, all columns that sum to 3 in the design belong to the group of the intercept column. Also, every column that sums to 3 must have inner products equal to $-1$ with every other remaining factor column in the design that sums to $-1$ as they do not belong to the same group.

% It is important to note that every row or column in $\mathbf{K}_{(N,p,s)}$ describes the inner product of a particular column in the design with every other column in the design. For example, the second row (or column) describes the inner products of the second column of the design with every other column. 
When $p$ is divisible by $s$, every row or column in $\mathbf{K}_{(N,p,s)}$ has the same number of entries for the value of 3 (when $p>s$) and $-1$. This means that regardless of the position of the intercept column in the design, the numbers of columns that must sum to 3 and those that must sum to $-1$ do not change. 

When $p$ is not divisible by $s$, there exist groups of unequal sizes. Consider the example matrix $\mathbf{K}_{(15,14,4)}$ given in Equation (\ref{eq:15_14_4}). For a design with such an information matrix, if the intercept column corresponds with the first column and row of the Ehlich matrix, the design must have two factor columns that sum to $3$ and eleven factor columns that sum to $-1$. However, if the intercept column corresponds with the last row and column of the Ehlich matrix, the design must have three factor columns that sum to $3$ and ten factor columns that sum to $-1$. Therefore, when $p$ is not divisible by $s$, there are two sets of design solutions that differ in the number of factor columns that sum to 3 and $-1$ by one, while other constraints described above remain the same.

\subsubsection{Initial two-factor design} \label{ssec:m3_start}

Our enumeration starts with a two-factor design whose information matrix equals
\spacingset{1}
\begin{equation}
\mathbf{K}_{(N,3,3)} =
\begin{bmatrix}
\phantom{0}N & -1 & -1\\
-1 & \phantom{0}N & -1\\
-1 & -1 & \phantom{0}N\\
\end{bmatrix}. \label{eq:starting_design}
\end{equation}
\spacingset{1.9}
This design has one column for the intercept and two factor columns that sum to $-1$, and whose inner product equals $-1$ too. The two factor columns of the design involve $(N+1)/4$ replicates for each of the design points $(-1,-1)$, $(-1,1)$ and $(1,-1)$, and $(N-3)/4$ replicates for the design point $(1,1)$.

The initial design's information matrix is obviously an Ehlich matrix with $s=3$, $r=1$, $u=3$ and $v=0$. Adding columns to that design can only produce designs with an information matrix of the form $\mathbf{K}_{(N,p,s)}$ where $s \geq 3$. This is not a problem, because designs with $s \le 2$ are neither D- nor A-optimal when $N \geq 7$ and $p \geq 4$ (\citet{galil1980d,sathe199912,king2020direct}). Therefore, it is not interesting to enumerate designs for these cases.

\subsubsection{Extension}

Our enumeration of $p$-column designs with an information matrix of the form $\mathbf{K}_{(N,p,s)}$ starts from the three-column design with an information matrix $\mathbf{K}_{(N,3,3)}$. The enumeration involves a total of $p-3$ extensions with one extra column. These $p-3$ one-column extensions take place in several phases. The number of phases is $r$ in the event $p$ is divisible by $s$, and $r+1$ otherwise. 

That an $(r+1)$st phase is required in the event $p$ is not divisible by $s$ is due to a technical issue. The technical issue is that one of the columns in the model matrix {\bf X} corresponds to the intercept in the main-effects model, and therefore is a column of ones. In the event $p$ is divisible by $s$, we can assume without loss of generality that the first column of {\bf X} is the intercept column. Automatically, the first row and column of the information matrix $\mathbf{K}_{(N,p,s)}$ correspond then also to the intercept. The intercept column then belongs to the first of the $s$ groups of columns of the design to be constructed. However, in the event $p$ is not divisible by $s$, then, inevitably, $v>0$ and at least one of the $s$ groups of columns involves $r+1$ rather than $r$ columns. In that event, we can no longer assume without loss of generality that the intercept column belongs to the first group. Instead, we need to allow for two possibilities: one possibility is that the intercept column belongs to a group of $r$ columns, while the other possibility is that the intercept column belongs to a group of $r+1$ columns. Designs in which the intercept column belongs to a group of $r$ columns have one less column that sums to 3 than designs in which the intercept column belongs to a group of $r+1$ columns. These two types of designs are not isomorphic, so it is essential that our enumeration procedure creates the two types of designs in the event $p$ is not divisible by $s$. In the next few paragraphs, we start by describing Phases 1 to $r$ of the extension, in which the technical issue is not yet relevant, and end by describing Phase $r+1$, in which it is.

% From Equation (\ref{form_m3}), it is clear that choosing appropriate subsets of columns from any design of the form $\mathbf{K}_{(N,p,s)}$, will result in designs of the form $\mathbf{K}_{(N,ts,s)}$ for all values $t$ where $1 \leq t \leq p/s$. To illustrate, c

Phase 1 of the extension procedure extends the initial design with an information matrix $\mathbf{K}_{(N,3,3)}$, which involves three groups of one column, where the first column is the intercept column and the other two columns sum to $-$1, to a design with $s$ groups of one column, where the first column is the intercept column and the remaining $s-1$ columns sum to $-$1, and inner products of $-1$ between any pair of columns that sum to $-1$. This requires the initial design to be extended a total of $s-3$ times. During phase 1, we sequentially obtain all non-isomorphic designs with information matrices $\mathbf{K}_{(N,4,4)}, \mathbf{K}_{(N,5,5)},\dots, \mathbf{K}_{(N,s,s)}$. In the simplest enumeration case where $s=p$ (and therefore $r=1$, $u=s$ and $v=0$), our algorithm stops here because that case only requires $s-1$ columns that sum to $-1$ and have inner products of $-1$ (see Section~\ref{sub:all_constraints}). 

In all cases where $s<p$, the enumeration procedure requires $r-1$ additional phases. In order to understand the goals of these additional phases, consider first the enumeration of D- and A-optimal designs with 15 runs and 13 factors. These designs have an information matrix equal to $\mathbf{K}_{(15,14,4)}$ in Equation~(\ref{eq:15_14_4}), implying that $s=4$, $r=3$, $u=2$ and $v=2$. This implies that the designs involve four groups of columns, that two of these groups have three columns, and that the two remaining groups have four columns. The enumeration of the designs proceeds as follows:
\begin{itemize}
\item The enumeration procedure starts from the initial design with the information matrix $\mathbf{K}_{(N,3,3)}$ given in Equation~(\ref{eq:15_14_4}) with $N=15$. That design has $s=p=3$, meaning that it has three groups of columns only, and each group consists of one column, one of which is the intercept column. Our enumeration procedure assumes that the three columns of the initial design correspond to the first column of the first group of factors required for the 13-factor design, the first column of the second group and the first column of the third group. These columns correspond to rows and columns 1, 4 and 7 in $\mathbf{K}_{(15,14,4)}$.
\item Since the required 13-factor designs require $s=4$ groups of columns, Phase 1 of our enumeration procedure must add a fourth column to the initial design. At this point in the algorithm, this column is the only column in the fourth group of factors and corresponds to column 11 in $\mathbf{K}_{(15,14,4)}$. At this point, we have constructed all designs with an information matrix of the form $\mathbf{K}_{(N,4,4)}$, the first two groups of columns need two additional columns, while the last two groups need three additional columns.
\item Phase 2 of the enumeration procedure adds four more columns to the design, one column per group. The first three columns added correspond to rows and columns 5, 8 and 12 in $\mathbf{K}_{(15,14,4)}$, and are assigned to the second, third and fourth group. Each of these columns sums to $-1$, has an inner product of 3 with the other column in the same group, and has inner products of $-1$ with the columns in the other groups. The fourth column added is assigned to group 1, sums to 3, and has inner products of $-1$ with the columns in the other groups. This column corresponds to row and column 2 in $\mathbf{K}_{(15,14,4)}$. The designs produced by Phase 2 have information matrices of the form $\mathbf{K}_{(15,5,4)}$, $\mathbf{K}_{(15,6,4)}$, $\mathbf{K}_{(15,7,4)}$ and $\mathbf{K}_{(15,8,4)}$.
\item Phase 3 of the enumeration procedure also adds four columns to the designs outputted by Phase 2, again one column per group. The first three columns added correspond to rows and columns 6, 9 and 13 in $\mathbf{K}_{(15,14,4)}$, and are assigned to the second, third and fourth group. Each of these columns sums to $-1$, has an inner product of 3 with the other two columns in the same group, and has inner products of $-1$ with the columns in the other groups. The fourth column added in Phase 3 is assigned to group 1, sums to 3, has an inner product of 3 with the single column that sums to 3 previously added in group 1 in Phase 2, and has inner products of $-1$ with the columns in the other groups. This column corresponds to row and column 3 in $\mathbf{K}_{(15,14,4)}$. The designs produced by Phase 3 have information matrices of the form $\mathbf{K}_{(15,9,4)}$, $\mathbf{K}_{(15,10,4)}$, $\mathbf{K}_{(15,11,4)}$ and $\mathbf{K}_{(15,12,4)}$.
\item After Phase 3, groups 1--4 all have three columns, meaning that two of the groups are complete and two other groups still lack a fourth column. Due to the technical issue involving the intercept, Phase 4 adds a fourth columns to the two other groups in two different ways:
\begin{itemize}
\item The first way is that a fourth column is added to the group containing the intercept column as well as to a group not containing the intercept column. The column added to the group contaning the intercept column sums to 3, while the other column added to the group not containing the intercept sums to $-$1. Both added columns have an inner product of 3 with the other three columns in their group, and have inner products of $-1$ with all columns in the other groups. The final design then has one group of columns consisting of the intercept column and three columns that sum to 3. The 10 remaining columns then all sum to $-1$. For the information matrix to be of the standard Ehlich form in Equation~\ref{eq:15_14_4}, we have to assume here that the added columns appear in groups 3 and 4, and that one of these two groups contains the intercept column. 
\item The second way is that a fourth column is added to two groups not containing the intercept column. Both added columns then sum to $-1$, have an inner product of 3 with the other three columns in their group, and have inner products of $-1$ with all columns in the other groups. The final design then has one group of columns consisting of the intercept column and two columns that sum to 3. The 11 remaining columns then all sum to $-1$. For the information matrix to be of the standard Ehlich form given in Equation~\ref{eq:15_14_4}, we have to assume here that the added columns appear in groups 3 and 4, and that neither of these two groups contains the intercept column. 
\end{itemize}
%Phase 4 of the algorithm therefore only adds a column to group 3 and a column to group 4. The added columns sum to $-1$, have an inner product of 3 with the other three columns in their group, and have inner products of $-1$ with all columns in the other groups. This result in designs with information matrices of the form $\mathbf{K}_{(15,13,4)}$, $\mathbf{K}_{(15,14,4)}$. Note that Phase 4 is only required because designs because $p$ is not divisble by $s$, which implies that $v>0$ and that not all of the $s$ groups involve the same number of columns.
\end{itemize}

This example illustrates that Phases 2 to $r$ of our enumeration procedure extend the designs with $s$ groups of one column to designs with $s$ groups of $r$ columns. These $r-1$ additional phases are needed whenever $s<p$. Each of the phases adds one column to each of the $s$ groups. The columns added to groups $2,\dots,s$ should sum to $-1$, have inner products of $3$ with the columns already present in their group, and have inner products of $-1$ with all columns in the other groups. The column added to group 1 should sum to 3, have inner products of 3 with any columns already present in group 1, and have inner products of $-1$ with all columns in groups $2,\dots,s$. 

The input to Phase 2 of the enumeration procedure is the set of all non-isomorphic designs with an information matrix of the form $\mathbf{K}_{(N,s,s)}$. Its output is $s$-fold. More specifically, Phase 2 outputs all non-isomorphic designs with information matrices of the forms $\mathbf{K}_{(N,s+1,s)},\dots,\mathbf{K}_{(N,2s,s)}$. Phase 3 is similar to Phase 2, in that it generates all non-isomorphic designs with information matrices of the forms $\mathbf{K}_{(N,2s+1,s)},\dots,\mathbf{K}_{(N,3s,s)}$, starting from the designs with information matrix $\mathbf{K}_{(N,2s,s)}$ outputted by Phase 2. In general, for any $i \in \{2,\dots,r\}$, Phase $i$ of the enumeration creates all non-isomorphic designs with $\mathbf{K}_{(N,(i-1)s+1,s)},\dots,\mathbf{K}_{(N,is,s)}$, starting from the designs with information matrix $\mathbf{K}_{(N,(i-1)s,s)}$ outputted by Phase $i-1$. 

In the event $p$ is divisible by $s$, Phase $r$ is the final phase of the extension procedure. Otherwise, we need an additional phase, Phase $r+1$. This is because, when $p$ is not divisible by $s$, $v>0$, which means that at least one of the $s$ groups of columns involves $r+1$ instead of $r$ columns. Phase $r+1$ of the extension procedure adds an $(r+1)$st column to each of the $v$ groups that require $r+1$ columns. Phase $r+1$ is more involved than the previous phases because a design with an information matrix of the form $\mathbf{K}_{(N,p,s)}$ can then be of two types. This is due to the technical issue involving the intercept column. 

The first type of design is obtained by extending the designs with information matrices of the form $\mathbf{K}_{(N,rs,s)}$ obtained at the end of Phase $r$ with $v$ columns that sum to $-1$. The $v$ added columns are assigned to the last $v$ of the $s$ groups of columns in the design, neither of which contains the intercept column. The added columns have inner products of 3 with the other columns in their group, and inner products of $-1$ with all other columns.

The second type of design is obtained by extending the designs with information matrices of the form $\mathbf{K}_{(N,rs,s)}$ obtained at the end of Phase $r$ with one column that sums to 3 and $v-1$ columns that sum to $-1$. The $v$ added columns are again assigned to the last $v$ of the $s$ groups of columns in the design, one of which is assumed to contain the intercept column. The added columns obviously have inner products of 3 with the other columns in their group, and products of $-1$ with all other columns.

\subsection{Possible sets of columns for extension} \label{ssec:m3_setcolumns}

% One important features of all matrices $\mathbf{X}$ that are of the form $\mathbf{K}_{(N,p,s)}$ and that have its first column as the intercept column is that all columns can be divided in $s$ groups, such that the inner product is equal to $-1$ between all pairs of columns that belong to different groups and the inner product is equal to 3 between all pairs of columns that belong to same group. These groups can be identified by simply taking the indices of the rows (or columns) in the matrix $\mathbf{K}_{(N,p,s)}$ that correspond to the $s$ different \textit{blocks}.

To construct all designs with an information matrix of the form $\mathbf{K}_{(N,p,s)}$, we need sets of columns that sum to $-1$ and $3$. For a given run size $N$, there are $N \choose (N+1)/2$ columns that sum to $-1$ and $N \choose (N+3)/2$ columns that sum to 3. We refer to the set of columns that sum to $-1$ by $\zeta_{N}^{-1}$ and to the set of columns that sum to 3 by $\zeta_{N}^{3}$. The matrix $\mathbf{K}_{(N,p,s)}$ defines what the inner products between all pairs of columns in the design matrix $\mathbf{X}$ must be. Due to this structure and our choice to start with a design that has an information matrix of the form $\mathbf{K}_{(N,3,3)}$, not every column from the sets $\zeta_{N}^{-1}$ and $\zeta_{N}^{3}$ is eligible for addition to the initial two-factor design. 

More specifically, eligible columns from $\zeta_{N}^{3}$ must have an inner product of $-1$ with the two factor columns of the initial design (see Section \ref{sub:all_constraints}). This reduces the set of candidate columns substantially. We refer to the subset of $\zeta_{N}^{3}$ containing only the eligible columns as $\zeta_{N}^{3*}$. Combinatorially, for a given $N$, the number of columns in $\zeta_{N}^{3*}$ can be expressed as 
\begin{equation*}
    \sum_{i=0}^{x} {x \choose i}{x+1 \choose i+1}{x+1 \choose i}^{2},
\end{equation*}

\noindent where $x = (N-3)/4$. As an example, for $N=19$, the number columns in $\zeta_{N}^{3}$ is 75582, whereas $\zeta_{N}^{3*}$ includes only 9030 columns. 

Since the two factor columns in the initial design have an inner product of $-1$, they belong to different groups. As a result, the  eligible columns in $\zeta_{N}^{-1}$ depend on their relation with these factor columns, as follows:
\begin{itemize}
    \item Columns that belong to a group that does not contain the first and second factor column of the initial design have an inner product of $-$1 with both the these factor columns. We denote this set of columns with $\zeta_{N}^{-1*(s)}$.
    \item Columns that belong to the group with the first factor column of the initial design have inner products of 3 and $-$1, respectively, with the first and second factor column of the initial design.
    \item Columns that belong to the group with the second factor column of the initial design have an inner product of $-$1 and 3, respectively, with the first and second factor column of the initial design.
\end{itemize}

We denote the subset of $\zeta_{N}^{-1}$ for any of the above three cases as $\zeta_{N}^{-1*}$. Combinatorially, for a given $N$, the number of columns in $\zeta_{N}^{-1*}$ equals 
\begin{equation*}
    \sum_{i=0}^{x} {x \choose i}{x+1 \choose i+1}\bigg[{x+1 \choose i+1}^{2} + 2{x+1 \choose i}^{2}\bigg],
\end{equation*}
while the number of columns in $\zeta_{N}^{-1*(s)}$ equals
\begin{equation*}
   \sum_{i=0}^{x} {x \choose i} {x+1 \choose  i+1}^{3}.
\end{equation*}
\noindent For most phases of our enumeration, we require $\zeta_{N}^{3*}$ and $\zeta_{N}^{-1*}$, while for the initial phase, we need $\zeta_{N}^{-1*(s)}$.

The reduction in numbers of columns from $\zeta_{N}^{-1}$ to $\zeta_{N}^{-1*}$ and $\zeta_{N}^{-1*(s)}$is substantial. For example, with $N=19$, the number of columns in $\zeta_{N}^{-1}$ is 92378, whereas it is only 28686 in $\zeta_{N}^{-1*}$ and 10626 in $\zeta_{N}^{-1*(s)}$. 

\subsection{Isomorphism testing} \label{ssec:m3_esteps}

%For a given run size $N$, number of parameters $p$, and values for $s,r,u$, and $v$, where $p,s \geq 3$, our enumeration proceeds as follows:

%\begin{itemize}[noitemsep,topsep=0pt]
%    \item[Step 1:] Generate the two-factor starting design from Section \ref{ssec:m3_start}.
 %   \item[Step 2:] Generate the three sets of candidate columns: $\zeta_{N}^{3*}, \zeta_{N}^{-1*}$ and $\zeta_{N}^{-1*(s)}$ of Section \ref{ssec:m3_setcolumns}.
    % \begin{itemize}[noitemsep,topsep=0pt]
    %     \item \textbf{Step 2.1:} Generate the set $\zeta_{N}^{3}$ containing all possible columns that sum to 3. Then, reduce this set by only keeping columns that have an inner product of $-1$ with both the first and second factor column of the starting design from Step 1. This produces the first set of candidate columns, $\zeta_{N}^{3*}$, to be considered in Step 3.
    %     \item \textbf{Step 2.2:} Generate the set $\zeta_{N}^{-1}$ containing all possible columns that sum to $-1$. Then, reduce this set by only keeping columns whose inner products with the first and second factor column of the starting design are (i) $-1$ and $-1$, (ii) $3$ and $-1$, or (iii) $-1$ and $3$, respectively. This produces the second set of candidate columns, $\zeta_{N}^{-1*}$, to be considered in Step 3.
    %     \item \textbf{Step 2.3:} From the set $\zeta_{N}^{-1*}$, retain columns that have an inner product of $-1$ with both columns of the starting design. This produces the third set of candidate columns, $\zeta_{N}^{-1*(s)}$, to be considered in Step 3. 
    % \end{itemize}
In most of the cases, each successful one-column extension results in multiple designs. Some of them can be converted to each other by row permutations, column permutations or switching the signs of a column. When this is the case, the designs belong to the same isomorphism class. As isomorphic designs have identical statistical properties, it makes sense to retain only one per isomorphism class. For this purpose, we convert the designs obtained to their Nauty minimal form \citep{mckay2014practical} using the OA package \citep{eendebak2019oapackage}. This conversion produces unique representatives for each isomorphism class called canonical forms. The canonical forms may have initial pairs of columns different from the columns of the two-factor initial design. Two databases are therefore utilized at each one-column extension stage. The first database stores all canonical forms, while the second database stores the original extended designs associated with each of the canonical forms in the first database. In subsequent extension stages, we always extend the designs in the second database as they retain the first two columns of the starting design. This is necessary to justify using the reduced sets  $\zeta_{N}^{3*}, \zeta_{N}^{-1*}$ and $\zeta_{N}^{-1*(s)}$ of candidate columns. 

Each one-column extension starts with all the non-isomorphic designs with the same values of $N,s,r,u$ an $v$. At the start of the extension, both databases are empty. Next, the first extended design is converted to its canonical form. That form is stored in the first database, and the design itself in the second database. For each new design, a check is performed whether its canonical form is already in the first database. If this is the case, the design is discarded. If this is not the case, both the first and second database are updated with a new canonical form and a new design, respectively.

\section{Results} \label{sec:enu_results}

In this section, we detail the results of our complete enumeration. In Section \ref{ssec:enu_designs}, we give the numbers of non-isomorphic designs we obtained and in Section \ref{ssec:char_results}, we characterize the designs and list the best of the designs.

\subsection{Enumeration} \label{ssec:enu_designs}

We used our algorithm to generate all non-isomorphic designs with information matrices of the form $\mathbf{K}_{(N,p,s)}$ for $s \geq 3$ (see Section \ref{ssec:m3_start}) and $N$ = 7, 11 and 15. For $N = 19$, the computations required were much more demanding, and hence we focused on the forms $\mathbf{K}_{(19,p,s)}$ likely to yield D- and A-optimal designs. In this article, we present the enumeration results for $N=15$, while the results for the other run sizes are provided in the online supplementary material.

% due to a much greater need for computations, we generate $\mathbf{X}$ matrices for only a limited set of forms most of which correspond to D- and/or A-optimal designs, and not all non-isomorphic $\mathbf{X}$ matrices for all existing forms $\mathbf{K}_{(19,p,s)}$, like we did with $N$ = 7, 11 and 15. To generate designs with 19 runs, we found our original procedure described in Section \ref{sec:algorithm} too slow. To tackle the case with $N=19$, we adapted the original algorithm on a case-by-case basis by adding additional constraints to minimize the number of columns in the candidate set of columns.

% \subsubsection{$N$ = 15}
In Figure \ref{tab:enu_15}, we give the total numbers of non-isomorphic designs with information matrices of the form $\mathbf{K}_{(15,p,s)}$ for different combinations of $p$ and $s$ where $4 \leq p \leq N$ and $s \geq 3$. A cell corresponding to a given pair $(p,s)$ gives the number of non-isomorphic designs for the form $\mathbf{K}_{(15,p,s)}$. If the cell is bordered by a dark line, then the corresponding number of designs are D-optimal, while if the cell is shaded, they are A-optimal. We provide a bold font for numbers that are the largest in each column. For cases where $p$ is not divisible by $s$, we give the sum of the number of designs for both type-1 and type-2.

\spacingset{1}
\begin{table}[ht]
  \centering
  \footnotesize
  \addtolength{\tabcolsep}{-0.2em}
    \begin{tabular}{cccccccccccccc}
    \multirow{13}[1]{*}{($s$)} & \multicolumn{1}{r|}{15} &       &       &       &       &       &       &       &       &       &       &       & \textbf{10} \\
          & \multicolumn{1}{r|}{14} &       &       &       &       &       &       &       &       &       &       & 18 & 0 \\
          & \multicolumn{1}{r|}{13} &       &       &       &       &       &       &       &       &       & 45 & 0 & 0 \\
          & \multicolumn{1}{r|}{12} &       &       &       &       &       &       &       &       & 88 & 0 & 0 & 0 \\
          & \multicolumn{1}{r|}{11} &       &       &       &       &       &       &       & 137 & 0 & 0 & 0 & 0 \\
          & \multicolumn{1}{r|}{10} &       &       &       &       &       &       & \squared{171} & 52 & 0 & 0 & 0 & 0 \\
          & \multicolumn{1}{r|}{9} &       &       &       &       &       & \squared{166} & \squared{286} & 23 & 0 & 0 & 0 & 0 \\
          & \multicolumn{1}{r|}{8} &       &       &       &       & \squared{\cellcolor{blue!20}117} & 809 & 869 & \squared{191} & 46 & 8 & 0 & 0 \\
          & \multicolumn{1}{r|}{7} &       &       &       & \squared{\cellcolor{blue!20}54} & \cellcolor{blue!20}979 & 3807 & 2464 & 407 & \squared{13} & 0 & 0 & 0 \\
          & \multicolumn{1}{r|}{6} &       &       & \squared{\cellcolor{blue!20}20} & 500 & 5298 & \cellcolor{blue!20}12277 & 6690 & 1322 & \squared{137} & 0 & 0 & 0 \\
          & \multicolumn{1}{r|}{5} &       & \squared{\cellcolor{blue!20}8} & 144 & 2107 & \textbf{10974} & 12701 & \cellcolor{blue!20}2857 & 4840\cellcolor{blue!20} & \cellcolor{blue!20}2188 & 0 & 0 & 0 \\
          & \multicolumn{1}{r|}{4} & \squared{\cellcolor{blue!20}4} & 30 & \textbf{345} & \textbf{2166} & 3442 & \textbf{27527} & \textbf{60699} & \textbf{44937} & \textbf{8356} & \textbf{2265} & \cellcolor{blue!20}\textbf{158} & \squared{\cellcolor{blue!20}3} \\
          & \multicolumn{1}{r|}{3} & \textbf{8} & \textbf{35} & 118 & 1802 & 6273 & 4232 & 14762 & 12451 & 2219 & 458 & 0 & 0 \\
\cmidrule{3-14}    \multicolumn{2}{c}{\multirow{2}[1]{*}{}} & 4 & 5 & 6 & 7 & 8 & 9 & 10 & 11 & 12 & 13 & 14 & 15 \\
    \multicolumn{2}{c}{} & \multicolumn{12}{c}{($p$)} \\
    \end{tabular}%
    \captionof{figure}{Numbers of non-isomorphic 15-run designs with an information matrix of the form $\mathbf{K}_{(15,p,s)}$ for different numbers of parameters $p$ and numbers of blocks $s$. In each column, the bordered and shaded cells indicate the numbers of designs that are D- and A-optimal, respectively, while the bold font indicate the numbers that are the largest.} 
  \label{tab:enu_15}%
  \end{table}
\spacingset{1.9}

As $p$ approaches $N$, there are many values of $s$ without solution for $\mathbf{X}$. For values of $p$, where the values for $s$ associated with D-optimal designs are not the same as those associated with A-optimal designs, the latter designs have a smaller value for $s$. For these combinations, there are more A-optimal designs than there are D-optimal designs.

For two out of the twelve values of $p$, we are unable to provide complete catalogs of D- and A-optimal designs. These are:
\begin{itemize}
    \item $p=13$: \citet{gk_opt_wei_des} showed that both forms $\mathbf{K}_{(15,13,6)}$ and $\mathbf{K}_{(15,13,7)}$, which correspond to the candidate information matrices for a D-optimal design, do not have a solution for $\mathbf{X}$. The form $\mathbf{K}_{(15,13,5)}$, which corresponds to the candidate information matrix for an A-optimal design, also has no solution for $\mathbf{X}$ \citep{chadjiconstantinidis1996optimal}. These authors also give the candidate information matrix for the global A-optimal design, which is not of the form given in Equation (\ref{form_m3}) and does have a solution for $\mathbf{X}$. Of all forms $\mathbf{K}_{(15,13,s)}$, the form $\mathbf{K}_{(15,13,4)}$ is the one with the lowest value for the trace of $(\mathbf{X'X})^{-1}$, and has a relative A-efficiency of 99.88\% with respect to the A-optimal candidate form. Also, the form $\mathbf{K}_{(15,13,8)}$ has the largest value for the determinant of all designs we obtained, and has a relative D-efficiency of 99.90\% with respect to the D-optimal candidate forms ($\mathbf{K}_{(15,13,6)}$ and $\mathbf{K}_{(15,13,7)}$) which do not have a solution for $\mathbf{X}$. We conjecture that the form $\mathbf{K}_{(15,13,8)}$ is the feasible candidate form of the information matrix for a globally D-optimal design.
    \item $p=14$: Here, $\mathbf{K}_{(15,14,6)}$, which is the candidate for a D-optimal design and $\mathbf{K}_{(15,14,5)}$, which is the candidate for an A-optimal design, do not have a solution for $\mathbf{X}$ as shown in \citet{gk_opt_wei_des} and \citet{chadjiconstantinidis1996optimal}, respectively. The latter authors further prove that the form $\mathbf{K}_{(15,14,4)}$ is the true candidate form for an A-optimal design. We were able to generate all non-isomorphic designs for this form. This form has a relative D-efficiency of 99.53\% with respect to the original candidate form ($\mathbf{K}_{(15,14,6)}$). We conjecture that the form $\mathbf{K}_{(15,14,4)}$ is also the feasible candidate form of the information matrix for a D-optimal design.
\end{itemize}

\noindent For $N=p=15$, the original candidate forms with the maximum determinant ($\mathbf{K}_{(15,15,6)}$) and minimum trace ($\mathbf{K}_{(15,15,5)}$) have no solution for $\mathbf{X}$ (\citet{gk_opt_wei_des} and \citet{chadjiconstantinidis1994exact}). However, \citet{orrick2005maximal} and \citet{chadjiconstantinidis1994exact} show that $\mathbf{K}_{(15,15,4)}$ is in fact the feasible candidate information matrix for a D- and A-optimal design. Therefore, the designs we obtained for this form are both D- and A-optimal.

\subsection*{Computing times}

In Figure \ref{tab:times_15}, we give the total computational time required to generate all non-isomorphic designs with information matrices of the form $\mathbf{K}_{(15,p,s)}$ for different combinations of $p$ and $s$ where $4 \leq p \leq N$ and $s \geq 3$. The layout of Figure \ref{tab:times_15} is similar to Figure \ref{tab:enu_15}, except that here instead of the numbers of non-isomorphic designs, we give the computational time required to generate the numbers of non-isomorphic designs given in Figure \ref{tab:enu_15}. We provide a bold font for computational times that are the largest in each column. For cases where $p$ is not divisible by $s$, the computational times in Figure \ref{tab:times_15} represent the total computational time required to generate all non-isomorphic designs of type-1 and type-2.

In general, we observe that the computational times required to generate designs with an information matrix of the form $\mathbf{K}_{(15,p,s)}$ with lower values for $s$ are larger than that for higher values for $s$.

\spacingset{1}
\begin{table}[!ht]
\footnotesize
  \centering
  \addtolength{\tabcolsep}{-0.3em}
    \begin{tabular}{cccccccccccccc}
    \multirow{13}[1]{*}{(s)} & \multicolumn{1}{r|}{15} &       &       &       &       &       &       &       &       &       &       &       & 11.02 \\
          & \multicolumn{1}{r|}{14} &       &       &       &       &       &       &       &       &       &       & 10.81 & - \\
          & \multicolumn{1}{r|}{13} &       &       &       &       &       &       &       &       &       & 10.39 & - & - \\
          & \multicolumn{1}{r|}{12} &       &       &       &       &       &       &       &       & 9.50 & - & - & - \\
          & \multicolumn{1}{r|}{11} &       &       &       &       &       &       &       & 8.07 & - & - & - & - \\
          & \multicolumn{1}{r|}{10} &       &       &       &       &       &       & \squared{6.09} & 19.93 & - & - & - & - \\
          & \multicolumn{1}{r|}{9} &       &       &       &       &       & \squared{4.20} & \squared{18.21} & 33.12 & - & - & - & - \\
          & \multicolumn{1}{r|}{8} &       &       &       &       & \squared{\cellcolor{blue!20}2.59} & 13.23 & 63.81 & \squared{93.48} & 91.49 & 91.44 & - & - \\
          & \multicolumn{1}{r|}{7} &       &       &       & \squared{\cellcolor{blue!20}1.63} & \cellcolor{blue!20}6.94 & 68.99 & 229.46 & 270.90 & \squared{263.14} & - & - & - \\
          & \multicolumn{1}{r|}{6} &       &       & \squared{\cellcolor{blue!20}1.10} & 3.94 & 43.02 & \cellcolor{blue!20}277.54 & 600.09 & 620.27 & \squared{595.83} & - & - & - \\
          & \multicolumn{1}{r|}{5} &       & \squared{\cellcolor{blue!20}0.69} & 2.30 & 15.03 & 107.12 & \textbf{337.75} & \cellcolor{blue!20}302.20 & 503.15\cellcolor{blue!20} & \cellcolor{blue!20}726.25 & - & - & - \\
          & \multicolumn{1}{r|}{4} & \squared{\cellcolor{blue!20}0.21} & 1.41 & \textbf{4.65} & \textbf{22.73} & 27.91 & 305.98 & \textbf{1895.52} & \textbf{3640.84} & \textbf{3446.84} & \textbf{4329.74} & \cellcolor{blue!20}\textbf{4213.43} & \squared{\cellcolor{blue!20}\textbf{4203.74}} \\
          & \multicolumn{1}{r|}{3} & \textbf{0.71} & \textbf{2.17} & 2.56 & 18.06 & \textbf{128.81} & 128.98 & 477.55 & 1105.38 & 1041.20 & 1212.15 & - & - \\
\cmidrule{3-14}    \multicolumn{2}{c}{\multirow{2}[1]{*}{}} & 4 & 5 & 6 & 7 & 8 & 9 & 10 & 11 & 12 & 13 & 14 & 15 \\
    \multicolumn{2}{c}{} & \multicolumn{12}{c}{(p)} \\
    \end{tabular}%

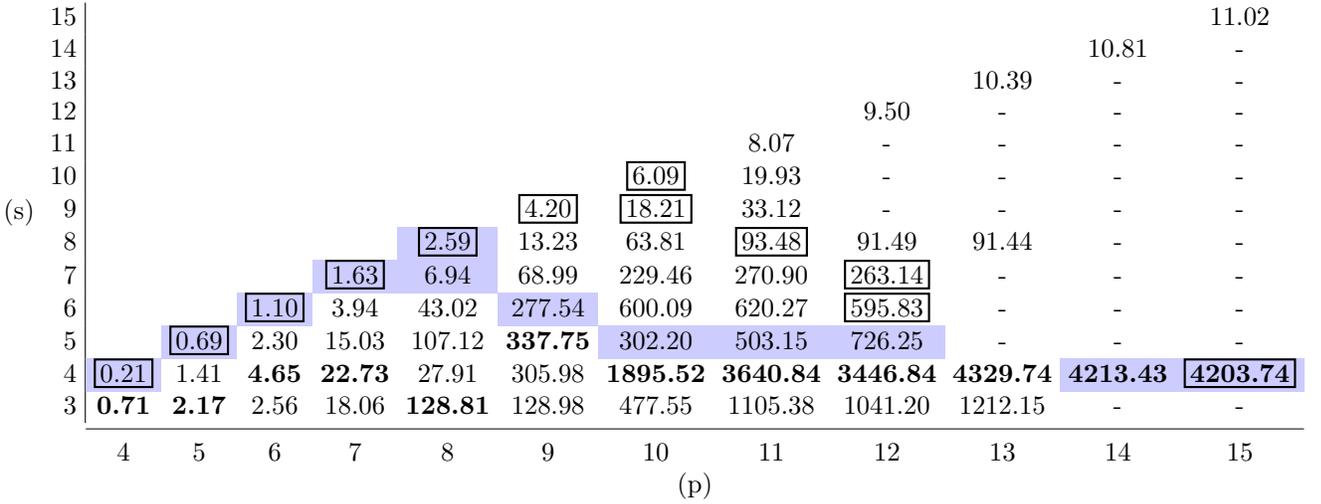
\captionof{figure}{The computation time required to generate all non-isomorphic 15-run designs with an information matrix of the form $\mathbf{K}_{(15,p,s)}$ for different numbers of parameters $p$ and numbers of blocks $s$. In each column, the bordered and shaded cells indicate the computation time for the forms of the information matrix that correspond to D- and A-optimal designs, respectively, while the bold font indicate the numbers that are the largest in each column.} 
  \label{tab:times_15}%
  \end{table}
\spacingset{1.9}

\subsection{Characterization} \label{ssec:char_results}

In order to rank our designs, we use the G$_2$-aberration criterion proposed by \citet{deng1999minimum}. In Section \ref{sssec:g2}, we introduce this criterion and in Section \ref{sssec:char_g2_results}, we characterize the best designs we obtained.

\subsubsection{G$_2$-aberration} \label{sssec:g2}

The G$_2$-aberration criterion is based on alias matrices that quantify the bias in main-effects estimates in the presence of active interaction effects. Denote by $\mathbf{X}_m$ the model matrix for the main-effects model including the intercept. If there are no active interaction effects, then the expected value of the response $\mathbf{Y}$ under the main-effects model is $\mathbf{X}_m \bm{\beta}_m$, where the first entry in $\bm{\beta}_m$ is the intercept and the remaining $k$ entries correspond to the factors' main effects. When there are active interaction effects of order $i$, this expected value becomes $\mathbf{X}_m \bm{\beta}_m + \mathbf{X}_i\bm{\beta}_i,$ where $\mathbf{X}_i$ is the matrix with all $i$-th order interaction contrast columns and $\bm{\beta}_i$ is the vector with all $i$-th order interaction effects. The matrix $\mathbf{A}_i=(\mathbf{X}_m'\mathbf{X}_m)^{-1}\mathbf{X}_m'\mathbf{X}_i$ is the alias matrix for the main-effects model associated with the $i$th-order interaction effects. The product $\mathbf{A}_i\bm{\beta}_i$ is the bias in the estimates of the intercept and the main effects when we fit the main-effects model ignoring the $i$th-order interaction effects. Since we generally care less about the bias in the estimate of the intercept, we ignore the first row of $\mathbf{A}_i$, and denote the matrix obtained by dropping that row by $\mathbf{A}^{*}_{i}$.

To minimize the overall bias in the estimates of the main effects due to the $i$th-order interaction effects, \citet{deng1999minimum} suggested to sequentially minimize the vector $(C_2, C_3, \dots, C_{k-1})$, where $C_i = tr({\mathbf{A}^{*}_{i} \mathbf{A}^{*}_{i}}').$ That vector quantifies the bias in the main-effect estimates in the presence of active interactions of orders 2, 3, \dots, $k-1$, respectively. Due to the hierarchy principle, which states that lower-order interactions are more likely to be active than higher-order interactions, the minimization of $C_2$ should therefore be prioritized over the minimization of $C_3$, the minimization of $C_3$ should be prioritized over the minimization of $C_4$, etc. This reasoning inspired \citet{deng1999minimum} to propose the G$_2$-aberration criterion: a design that sequentially minimizes the vector $(C_2, C_3, \dots, C_{k-1})$ performs best in terms of the G$_2$-aberration criterion and is called the minimum G$_2$-aberration design. Such a design minimizes the overall bias in the estimates of the main effects in the presence of higher-order interaction effects.

The quantities $C_2$ and $C_3$ can also be interpreted as measures of the overall aliasing between main effects and two-factor interactions and the overall aliasing among the two-factor interactions, respectively. In a similar fashion, the elements of the vector $(C_4, C_5, \dots, C_{k-1})$ can be interpreted as measures of aliasing of higher-order interaction effects. Due to the fact that higher-order interactions are generally unimportant, we only consider the $C_2$ and $C_3$ values when characterizing the non-isomorphic designs we enumerated.

In the literature, the G$_2$-aberration criterion has been mainly applied to orthogonal designs. These designs are D- and A-optimal for the main-effects model. So, in the literature, the G$_2$-aberration criterion can be viewed as a secondary criterion that can be used to distinguish between multiple designs that are all D- and A-optimal. The best designs according to this criterion are thus minimally aliased D- and A-optimal designs with respect to the main effects. This criterion has been used by other authors (see \citet{zhang2013minimum}, \citet{huda2013two} and \citet{hameed2025_m12} to rank non-orthogonal designs that are both D- and A-optimal for run sizes $N \equiv$ 1 (mod 4) and $N \equiv$ 2 (mod 4). In our paper, we also use the G$_2$-aberration criterion to rank our set of D- and A-optimal designs with run size $N \equiv$ 3 (mod 4). For these run sizes, for certain combinations of $N$ and $p$, the set of D-optimal designs can be different from the set of A-optimal designs. Therefore, for such cases, we report the minimally aliased D- and A-optimal design separately.

\subsubsection{Properties of the best designs} \label{sssec:char_g2_results}

The characterization results for designs with run size 15 is given in Figure \ref{tab:char_15}, while the results for the other run sizes are provided in the online supplementary material. The layout of Figure \ref{tab:char_15} is similar to Figure \ref{tab:enu_15}, except that here instead of the numbers of non-isomorphic designs, we give the lowest $C_2$ value observed for a design with an information matrix of the form $\mathbf{K}_{(15,p,s)}$ for which we obtained designs. In other words, we provide the $C_2$ value for the minimally aliased design for each form of the information matrix of the form $\mathbf{K}_{(15,p,s)}$. We provide a bold font for numbers that are the smallest in each column.

\spacingset{1}
\begin{table}[!ht]
\footnotesize
  \centering
  \addtolength{\tabcolsep}{-0.3em}
    \begin{tabular}{cccccccccccccc}
    \multirow{13}[1]{*}{(s)} & \multicolumn{1}{r|}{15} &       &       &       &       &       &       &       &       &       &       &       & 182.00 \\
          & \multicolumn{1}{r|}{14} &       &       &       &       &       &       &       &       &       &       & 105.00 & - \\
          & \multicolumn{1}{r|}{13} &       &       &       &       &       &       &       &       &       & 70.75 & - & - \\
          & \multicolumn{1}{r|}{12} &       &       &       &       &       &       &       &       & 49.06 & - & - & - \\
          & \multicolumn{1}{r|}{11} &       &       &       &       &       &       &       & 32.40 & - & - & - & - \\
          & \multicolumn{1}{r|}{10} &       &       &       &       &       &       & \squared{18.00} & 33.06 & - & - & - & - \\
          & \multicolumn{1}{r|}{9} &       &       &       &       &       & \squared{\textbf{4.57}} & \squared{\textbf{17.67}} & 32.83 & - & - & - & - \\
          & \multicolumn{1}{r|}{8} &       &       &       &       & \squared{\cellcolor{blue!20}\textbf{2.30}} & 8.67 & 20.72 & \squared{31.90} & 45.20 & 61.68 & - & - \\
          & \multicolumn{1}{r|}{7} &       &       &       & \squared{\cellcolor{blue!20}\textbf{1.11}} & \cellcolor{blue!20}4.53 & 11.78 & 20.97 & 31.74 & \squared{45.75} & - & - & - \\
          & \multicolumn{1}{r|}{6} &       &       & \squared{\cellcolor{blue!20}\textbf{0.50}} & 1.36 & 4.80 & \cellcolor{blue!20}11.96 & 20.50 & 31.27 & \squared{44.77} & - & - & - \\
          & \multicolumn{1}{r|}{5} &       & \squared{\cellcolor{blue!20}\textbf{0.20}} & 0.76 & 1.83 & 4.27 & 11.31 & \cellcolor{blue!20}20.28 & 29.54\cellcolor{blue!20} & \cellcolor{blue!20}43.46 & - & - & - \\
          & \multicolumn{1}{r|}{4} & \squared{\cellcolor{blue!20}\textbf{0.06}} & 0.41 & 1.06 & 2.71 & 5.52 & 11.57 & 18.42 & \textbf{26.98} & \textbf{40.31} & \textbf{58.00} & \cellcolor{blue!20}\textbf{78.53} & \squared{\cellcolor{blue!20}\textbf{100.75}} \\
          & \multicolumn{1}{r|}{3} & 0.18 & 0.61 & 1.60 & 3.78 & 7.15 & 13.12 & 20.81 & 30.44 & 44.39 & 60.10 & - & - \\
\cmidrule{3-14}    \multicolumn{2}{c}{\multirow{2}[1]{*}{}} & 4 & 5 & 6 & 7 & 8 & 9 & 10 & 11 & 12 & 13 & 14 & 15 \\
    \multicolumn{2}{c}{} & \multicolumn{12}{c}{(p)} \\
    \end{tabular}%
\captionof{figure}{The smallest value of $C_2$ obtained for 15-run designs with an information matrix of the form $\mathbf{K}_{(15,p,s)}$ for different numbers of parameters $p$ and numbers of blocks $s$. In each column, the bordered and shaded cells indicate the smallest values of $C_2$ for the forms of the information matrix that correspond to D- and A-optimal designs, respectively, while the bold font indicate the numbers that are the smallest overall.} 
  \label{tab:char_15}%
  \end{table}
\spacingset{1.9}

When $p=8$, the minimally aliased D- and A-optimal design with an information matrix of the form $\mathbf{K}_{(15,8,8)}$ has a $C_2$ value lower than the minimally aliased A-optimal design with an information matrix of the form $\mathbf{K}_{(15,8,7)}$. When $p=9$, the minimally aliased D-optimal design with an information matrix of the form $\mathbf{K}_{(15,9,9)}$ has a $C_2$ value lower than the minimally aliased A-optimal design with an information matrix of the form $\mathbf{K}_{(15,9,6)}$. When $p=10$, the minimally aliased D-optimal design with an information matrix of the form $\mathbf{K}_{(15,10,9)}$ has a $C_2$ value lower than both the minimally aliased D-optimal design with an information matrix of the form $\mathbf{K}_{(15,10,10)}$ and the minimally aliased A-optimal design with an information matrix of the form $\mathbf{K}_{(15,10,5)}$. When $p=11$, the minimally aliased A-optimal design with an information matrix of the form $\mathbf{K}_{(15,11,5)}$ has a $C_2$ value lower than the minimally aliased D-optimal design with an information matrix of the form $\mathbf{K}_{(15,11,8)}$. When $p=12$, the minimally aliased A-optimal design with an information matrix of the $\mathbf{K}_{(15,12,5)}$ has a $C_2$ value lower than both minimally aliased D-optimal designs with information matrices of the forms $\mathbf{K}_{(15,12,6)}$ and $\mathbf{K}_{(15,12,7)}$. We provide all the minimally aliased D- and A-optimal designs in the supplementary material to this paper.

Note that when $p \in \{ 11, 12\}$, minimally aliased designs with information matrices that do not correspond to D- and A-optimal designs have lower $C_2$ values than those that do. With reference to Table \ref{tab:char_15}, this result highlights that in certain cases, where $p$ approaches $N$, designs that are D- and A-optimal may have greater aliasing between main effects and interactions than designs that are not.

\section{Conclusion} \label{sec:conclusion}

The study of non-isomorphic D- and A-optimal main-effects designs has received much attention in the literature. Until now, existing studies have focused on complete enumerations of such designs, for run sizes $N$, where $N \equiv 0$ (mod 4), $N \equiv 1$ (mod 4), or $N \equiv 2$ (mod 4). For run sizes $N \equiv 3$ (mod 4), although the recent work of \citet{hameed2025_dropping} covers complete enumeration of D- and A-optimal designs, the number of cases covered by their work is limited. 

In our work, we propose an algorithm to construct all non-isomorphic designs with an information matrix of the complex form described by \citet{Ehlich1964}. These are the best known forms of information matrices corresponding to D- and A-optimal main-effects designs for run size $N \equiv 3$ (mod 4) for which (barring a few exceptional cases), it is possible to construct complete catalogs of non-isomorphic D- and A-optimal main-effects designs with information matrices belonging to the class of Ehlich matrices. 

Our algorithm can not only generate all designs for cases covered in \citet{hameed2025_dropping}, but also for all remaining cases not covered by their work. We use our algorithm to generate complete catalogs of D- and A-optimal main-effects designs for run sizes 7, 11, 15, and 19, find the designs that minimize aliasing between main-effects and two-factor interactions. In our paper, we discuss the enumeration and design characterization results for run size 15, while for the remaining run sizes 7, 11 and 15, we give these results in the online supplementary material along with all the designs discussed in the paper. As a result of our work, practitioners can now comfortably pick minimally aliased D- and A-optimal main-effects designs for any run size, including run sizes that are three more than a multiple of four.

\bibliographystyle{apalike}

\spacingset{1}
% \bibliography{short,bibliography.bib}
\bibliography{bibliography_m3.bib}
\spacingset{1.9}

\end{document}